\documentclass[prl,twocolumn,tightenlines,showpacs,nofootinbib]{revtex4}
\usepackage{bm,dcolumn,amsmath,graphicx,amsfonts,amssymb}
\usepackage{epsfig}

\begin{document}
\title{Parity non-conservation in rubidium atom}
 
\author{V. A. Dzuba, V. V. Flambaum, B. Roberts}
\affiliation{School of Physics, University of New South Wales,
Sydney, NSW 2052, Australia}

\date{ \today }

\begin{abstract}
Currently the theoretical uncertainty limits the interpretation of the
atomic parity non-conservation (PNC) measurements. We calculate the
PNC $5s$ - $6s$ electric dipole transition 
amplitude in rubidium and demonstrate that rubidium is a good
candidate to  search  for new physics beyond the standard
model since  accuracy of the atomic calculations 
in rubidium can be  higher than in cesium. PNC in cesium is currently
the best low-energy test of the standard model, therefore, similar
measurements for rubidium present a good 
option for further progress in the field. We also calculate
nuclear spin-dependent part of the parity non-conserving (PNC) amplitude which
is needed for the extraction of the nuclear anapole moment from the
PNC measurements.
\end{abstract}
\maketitle

\section{Introduction}

The study of parity non-conservation in atoms is a low-energy,
relatively inexpensive alternative to high-energy searches for new
physics beyond the standard model (see, e.g.~\cite{Ginges,DK04,DF12}).
For example, parity non-conservation in cesium is currently the best
low-energy test of the electroweak theory~\cite{Ginges,DF12}. 
It is due to high accuracy of the measurements~\cite{Wood} and its
interpretation~\cite{CsPNC} (see
also~\cite{CsPNCour,CsPNCold,Gingesrad}). The  
uncertainty of the measurements is 0.35\%~\cite{Wood} while the
uncertainty of the calculations is on the level of 0.4\% - 0.5\%
\cite{CsPNC,CsPNCour,CsPNCold}. This means that the interpretation of the
measurements is limited by the accuracy of atomic
calculations. The situation is similar for other atoms. For
example, accuracy of the PNC measurements for thallium is
1\%~\cite{Fortson} while accuracy of the calculations is 2.5 -
3\%~\cite{DzubaTl,KozlovTl}. It is believed that a good
option for further progress may come with the PNC measurements for
atoms or ions with electron structure similar to cesium but with higher
nuclear charge $Z$. Higher $Z$ would lead to larger PNC effect and
 would probably lead to better accuracy in the
measurements. The PNC measurements have been considered for the Ba$^+$
ion~\cite{BaII,Ba+} and are under progress for the Fr atom~\cite{FrPNC}
and Ra$^+$ ion~\cite{KVI}.
However, the accuracy of the calculations for these systems is unlikely
to be better than for cesium. Just on the contrary, higher $Z$ means
larger relativistic effects such as Breit and quantum electrodynamics
(QED) corrections, larger uncertainty due to the neutron skin effect,
etc. This would most 
likely lead to poorer accuracy of the calculations. Since the accuracy
of the calculations is a limiting factor even for cesium, it does make
sense in our view to look in opposite direction and to consider PNC in
rubidium. Rubidium is an alkali atom next to cesium but with smaller
$Z$. The same calculations as for cesium would lead to better accuracy
of the results for rubidium while the PNC amplitude in rubidium is
only seven times smaller than in cesium. Depending on the accuracy of
the measurements which can be achieved for rubidium, the study of the
PNC in this atom might be a good alternative for further progress in
the area. 

Rubidium was considered for anapole moment measurements in
Ref.~\cite{Orozco}. Corresponding atomic calculations were reported in
Ref.~\cite{Johnson,DF-hfs}. The calculations of the spin-independent
PNC amplitude of the $5s$ - $6s$ electric dipole transition in Rb were
performed in our early work~\cite{DzubaTl} with 2\% accuracy. Only
correlation corrections were considered while Breit, QED and other
small corrections were ignored. In this paper we perform  a more
detailed analysis of the PNC amplitude in Rb. This includes more
accurate treatment of the correlations and a detailed consideration of
the Breit, QED and neutron skin corrections. We demonstrate that the
analysis of the PNC effect in rubidium can be more reliable and
accurate than in cesium. This should encourage experimentalists to
consider the PNC measurements in rubidium. 

\section{Calculations}

The Hamiltonian describing parity-nonconserving electron-nucleus
interaction can be written as a sum of the nuclear-spin-independent (SI) and
the nuclear-spin-dependent (SD) parts (we use atomic units: $\hbar =
|e| = m_e = 1$): 
\begin{eqnarray}
     \hat H_{\rm PNC} &=& \hat H_{\rm SI} + \hat H_{\rm SD} \nonumber \\
      &=& \frac{G_F}{\sqrt{2}}                             
     \Bigl(-\frac{Q_W}{2} \gamma_5 + \frac{\varkappa}{I}
     {\bm \alpha} {\bm I} \Bigr) \rho({\bm r}),
\label{e1}
\end{eqnarray}
where  $G_F \approx 2.2225 \times 10^{-14}$ a.u. is the Fermi constant of
the weak interaction, $Q_W$ is the nuclear weak charge,
$\bm\alpha=\left(
\begin{array}
[c]{cc}%
0 & \bm\sigma\\
\bm\sigma & 0
\end{array}
\right)$ and $\gamma_5$ are the Dirac matrices, $\bm I$ is the
nuclear spin, and $\rho({\bf r})$ is the nuclear density normalized to 1.

Within the standard model
the weak nuclear charge $Q_W$ is given by~\cite{SM}
\begin{equation}
Q_W \approx -0.9877N + 0.0716Z.
\end{equation}
Here $N$ is the number of neutrons, $Z$ is the number of protons.

\begin{table}
\caption{Ionization energies (in cm$^{-1}$) and hyperfine structure constants
  $A$ (in MHz) for low states of $^{85}$Rb.}
\label{t:enhfs}
\begin{ruledtabular}
\begin{tabular}{lrrcc}
State & \multicolumn{2}{c}{Energies [cm$^{-1}$]} 
      & \multicolumn{2}{c}{$A$ [MHz]} \\ 
      & \multicolumn{1}{c}{Exper.\tablenotemark[1]} 
      & \multicolumn{1}{c}{Calc.} 
      & \multicolumn{1}{c}{Exper.} 
      & \multicolumn{1}{c}{Calc.} \\
\hline
$5s_{1/2}$ & 33691 & 33666 & 1011.9\tablenotemark[2] & 1016 \\
$5p_{1/2}$ & 21113 & 21145 &  120.5\tablenotemark[3] & 120.1 \\ 
$5p_{3/2}$ & 20874 & 20902 &         & \\
$4d_{3/2}$ & 14336 & 14362 &         & \\
$4d_{5/2}$ & 14335 & 14360 &         & \\
$6s_{1/2}$ & 13558 & 13509 & 239.18(3)\tablenotemark[4] & 239.2 \\
$6p_{1/2}$ &  9976 &  9973 & 39.11(3)\tablenotemark[5] & 38.87 \\  
$6p_{3/2}$ &  9898 &  9894 &         & \\
\end{tabular}
\tablenotetext[1]{Ref.~\cite{NIST}.} 
\tablenotetext[2]{Ref.~\cite{hfs5s}.}
\tablenotetext[3]{Ref.~\cite{hfs5p}.}
\tablenotetext[4]{Ref.~\cite{hfs6s}.}
\tablenotetext[5]{Ref.~\cite{hfs6p}.}
\end{ruledtabular}
\end{table}

\begin{table}
\caption{Electric dipole transition amplitudes (reduced matrix
  elements in a.u.) for low states of Rb.}
\label{t:e1}
\begin{ruledtabular}
\begin{tabular}{ccc}
\multicolumn{1}{c}{Transitions} &
\multicolumn{1}{c}{Exper.\tablenotemark[1]} &
\multicolumn{1}{c}{Calc.} \\ 
\hline
$5s_{1/2}$ - $5p_{1/2}$ & 4.231(3) & 4.246 \\ 
$5s_{1/2}$ - $5p_{3/2}$ & 5.977(4) & 5.994 \\
\end{tabular}
\tablenotetext[1]{Ref.~\cite{Rb-E1}.}
\end{ruledtabular}
\end{table}

To calculate the PNC amplitude we use the methods developed in our previous
works~\cite{CsPNCour,CPM}. The all-order correlation potential
$\hat \Sigma$~\cite{Sigma-all} is used to construct the so-called
Brueckner orbitals (BO) for the external electron. BO are found by solving
the Hartree-Fock-like equations with an extra operator $\hat \Sigma$: 
\begin{equation}
 (\hat H_0 +\hat \Sigma - \epsilon_a)\psi_a^{(\rm BO)}=0.
\label{eq:BO}
\end{equation}
Here $\hat H_0$ is the relativistic Hartree-Fock Hamiltonian, index
$a$ numerate valence states. The BO $\psi_a^{(\rm BO)}$ and energy
$\epsilon_a$ include dominating higher-order correlations.  
The parity non-conserving weak interaction as well as the electric
dipole interaction of the atom with laser light are included in the
framework of the time-dependent Hartree-Fock approximation~\cite{CPM}
which is equivalent to the well-known random-phase approximation
(RPA). 

In the RPA method, a single-electron wave function in external weak and
$E1$ fields is
\begin{equation}
\psi = \psi_0 + \delta\psi +X e^{-i\omega t}+Y e^{i\omega t} + \delta
Xe^{-i\omega t} + \delta Ye^{i\omega t},
\end{equation}
where $\psi_0$ is the unperturbed state, $\delta\psi$ is the
correction due to weak interaction acting alone, $X$ and $Y$ are
corrections due to the photon field acting alone, $\delta X$ and
$\delta Y$ are corrections due to both fields acting simultaneously,
and $\omega$ is the frequency of the PNC transition. The corrections
are found by solving  the system of RPA equations self-consistently 
for the core states
\begin{eqnarray}
&&(\hat H_0 - \epsilon_c)\delta\psi_c = - (\hat H_W + \delta \hat
V_W)\psi_{0c}, \nonumber \\ 
&&(\hat H_0 - \epsilon_c-\omega)X_c = - (\hat H_{E1} + \delta \hat
V_{E1})\psi_{0c}, \nonumber \\ 
&&(\hat H_0 - \epsilon_c+\omega)Y_c = - (\hat H_{E1}^{\dagger} + \delta \hat
V_{E1}^{\dagger})\psi_{0c}, \label{eq:RPA} \\ 
&&(\hat H_0 - \epsilon_c-\omega)\delta X_c = -\delta\hat V_{E1}\delta\psi_c -
\delta\hat V_WX_c - \delta\hat V_{E1W}\psi_{0c}, \nonumber \\ 
&&(\hat H_0 - \epsilon_c+\omega)\delta Y_c = -\delta\hat
V_{E1}^{\dagger}\delta\psi_c -
\delta\hat V_WY_c - \delta\hat V_{E1W}^{\dagger}\psi_{0c}, \nonumber 
\end{eqnarray}
where index $c$ numerates core states, $\hat H_W$ is either $\hat
H_{SI}$ or $\hat H_{SD}$ (see Eq.~\ref{e1}), $\delta \hat V_W$ and
$\delta \hat V_{E1}$ are corrections to the core potential due to the
weak and $E1$ interactions, respectively, and $\delta \hat V_{E1W}$ is
the correction to the core potential due to the simultaneous action of
the weak field and the electric field of the photon.

The PNC amplitude between valence states $a$ and $b$ in the RPA
approximation is given by
\begin{eqnarray}
&& E_{\rm PNC} = \langle \psi_b|\hat H_{E1} + 
 \delta\hat V_{E1}|\delta\psi_a\rangle + \nonumber \\
&& \langle \psi_b|\hat H_{W} + \delta\hat V_{W}|X_a\rangle +
\langle \psi_b|\delta\hat V_{E1W}|\psi_a\rangle = \nonumber \\
&&\langle \psi_b|\hat H_{E1} + \delta\hat V_{E1}|\delta\psi_a\rangle
+ \label{amp} \\ 
&& \langle \delta\psi_b|\hat H_{E1} + \delta\hat V_{E1}|\psi_a\rangle +
\langle \psi_b|\delta\hat V_{E1W}|\psi_a\rangle. \nonumber 
\end{eqnarray}
To include correlations in the calculation of the PNC amplitude one
needs to use BO for the valence states $a$ and $b$ in
(\ref{amp}). The corrections $\delta \psi_a$ and $\delta \psi_b$ to
BO $a$ and $b$ are also found with the use of the correlation
potential $\hat \Sigma$:
\begin{equation}
(\hat H_0 -\epsilon_a + \hat \Sigma)\delta\psi_a = -(\hat H_W +
\delta\hat V)\psi_{0a}. \label{eq:dpsiv}
\end{equation}
Note that the correlation potential $\hat \Sigma$ is the
energy-dependent operator. To calculate a BO and corrections to it one
should use the correlation potential at the energy of this state,
i.e. $\hat \Sigma \equiv \hat \Sigma(\epsilon_a)$ in (\ref{eq:BO}) and
(\ref{eq:dpsiv}). 

The way of calculation of the PNC amplitude described above does not
involve direct 
calculation of the electric dipole transition amplitudes or weak matrix
elements or even the energies, apart from the energies of the $5s$ and
$6s$ states. However, it is instructive to make comparisons with
available experimental data to have an idea of the accuracy of the
calculations. For this purpose we have performed the calculations of
the energies and magnetic dipole hyperfine structure constants of the
lowest $s$ and $p_{1/2}$ states of Rb as well as the electric dipole
transition amplitudes between these states. The calculations are done
with the use of the same approach and the same all-order operator
$\hat \Sigma$ as for the PNC calculations~\cite{Sigma-hfs}. The
results for the energies and the 
hyperfine structure are presented in Table~\ref{t:enhfs}, for E1
transition amplitudes in Table~\ref{t:e1}. Comparison with
experimental data shows that the accuracy of the calculations is about
0.1\% for the energies, 0.4 - 0.6\% for the hfs and about 0.3\% for
the E1 transition amplitudes. If we assume that the square root of the
product of hfs constants of $s$ and $p$ states can be used as a test
for the weak matrix elements ($\langle s|W|p\rangle \sim
\sqrt{A_sA_p}$) then the accuracy for the weak matrix elements is also
on the level of 0.3\%. Note however that the accuracy of this test is
limited. For example, the value of the ratio 
$\langle s|W|p\rangle/\sqrt{A_sA_p}$ is 4\% different in Hartree-Fock
and RPA approximations. This is because core polarization effects are
significantly different for weak and hfs interactions. Only $s$ and
$p_{1/2}$ states contribute to the core polarization for the weak
matrix elements. In the case of hfs interaction the $p_{3/2}$ states
also give a significant contribution. Since weak matrix elements are
simpler, the accuracy for them is expected to be higher than for the
hyperfine structure. 

\section{Results and discussion}

The  value of the spin-independent PNC amplitude for the $5s$ -
$6s$ transition in $^{87}$Rb (without  Breit, QED and neutron skin
corrections) is 
\begin{equation}
|E_{\rm PNC}| = 1.400 \times 10^{-12} ea_B (-Q_W/N).
\label{eq:epnc87}
\end{equation}
This is in very good agreement with the value
\[
|E_{\rm PNC}| = 1.39(2) \times 10^{-12} ea_B (-Q_W/N).
\]
presented in our early calculations~\cite{DzubaTl}.

Below we will discuss and compare different contributions to the
spin-independent PNC amplitudes in rubidium and cesium and point to
some advantages of using rubidium in searching for new physics beyond
the standard model. The contributions are presented in
Table~\ref{t:pnc}. We use the $^{87}$Rb isotope as an example.      

\begin{table}
\caption{Contributions to the parity non-conserving electric dipole
  transition $5s - 6s$ in $^{87}$Rb and $6s - 7s$ in $^{133}$Cs ($10^{-12}iea_B (-Q_W/N)$).}
\label{t:pnc}
\begin{ruledtabular}
\begin{tabular}{ldrdr}
Contribution &
\multicolumn{2}{c}{Rb} &
\multicolumn{2}{c}{Cs} \\
&\multicolumn{1}{c}{a.u.} &
\multicolumn{1}{c}{\%} &
\multicolumn{1}{c}{a.u.} &
\multicolumn{1}{c}{\%} \\
\hline
RPA          &  1.345 &   97\%& 8.899 & 99\%\\
Correlations &  0.054 &   4\% & 0.173 & 2\%\\
Subtotal     &  1.400 &  101\%     & 9.072 & 101\%\\
Breit        & -0.006 & -0.4\%&-0.055 & -0.6\%\\
QED          &    -0.003    &     -0.2\%  &-0.029 & -0.3\%\\
Neutron skin & -0.0008 & -0.06\% &-0.018 & -0.2\%\\
Total        &  1.390      &    100\%    & 8.970 & 100\% \\
\end{tabular}
\end{ruledtabular}
\end{table}

\paragraph{Correlations.}

The total correlation correction to the PNC amplitude is small for
both atoms. It is 2\% for cesium and 4\% for rubidium. The small value
of the correlation correction is the result of strong cancellation
between different terms. This is illustrated by the data in
Table~\ref{t:pnc1} where correlation corrections are presented for
each term in (\ref{amp}). In this table we use notation $\tilde d =
\hat H_{E1} + \delta\hat V_{E1}$ for short. The largest correlation
corrections are for those terms which have $\delta\psi$ weak
correction to the ground state. Corresponding corrections for Rb are
larger than for Cs. Indeed, the closer the valence
electron to the core the larger the correlation correction. The ionization
potential for Rb is larger than for Cs. This means that the valence
electron in Rb is closer to the core than in Cs. Also, the
cancellation between the contributions of 
$\delta\psi_a$ and  $\delta\psi_b$  in Rb is not as strong as in Cs.

Note that the strong cancellations between correlation corrections to
different terms in (\ref{amp}) do not mean poor numerical
accuracy. The theoretical uncertainty of the PNC amplitudes is mostly
due to missed terms while the numerical accuracy for all terms in
(\ref{amp}) is high. 

Table~\ref{t:pnc1} does not include non-Brueckner correlation
corrections, such as structure radiation, weak correlation potential
and renormalization of the wave functions~\cite{CsPNCour}. 
These corrections are suppressed by a small parameter
$E_{valence}/E_{core} \sim 1/10$ where  $E_{valence}$and $E_{core}$
are typical excitation energies of the valence and core
electrons. Moreover, their total 
contribution is practically zero for Cs~\cite{CsPNCour} due to
cancellation between different terms. It is expected to be very small for Rb
as well since all other relative contributions to $E_{PNC}$ in Rb and
Cs are  similar. 
 These terms can be calculated if progress is made with
the measurements. At the moment we just assume that they do not
contribute to the PNC amplitude or its uncertainty.

The data in Table~\ref{t:pnc1} show that the correlation correction to
the PNC amplitudes are similar in cesium and rubidium. Therefore,
similar uncertainty is expected. The uncertainty for cesium is 0.4 -
0.5\%~\cite{CsPNC,CsPNCour}. It is natural to expect the same
uncertainty for rubidium.

\begin{table}
\caption{Correlation corrections ($\Delta$) to the PNC amplitude, comparison
  between rubidium and cesium. Units: $10^{-12}iea_B (-Q_W/N)$.
Indexes $a$ and $b$ stand for $7s$ and $6s$ for Cs and for $6s$ and
$5s$ for Rb.}
\label{t:pnc1}
\begin{ruledtabular}
\begin{tabular}{ldddd}
Approxi-& 
\multicolumn{1}{c}{$\langle \delta\psi_a|\tilde d|\psi_b\rangle$} &
\multicolumn{1}{c}{$\langle \psi_a|\tilde d|\delta\psi_b\rangle$} &
\multicolumn{1}{c}{$\langle \psi_a|\delta \hat V_{E1W}|\psi_b\rangle$} &
\multicolumn{1}{c}{Total} \\
mation & & & & \\
\hline
\multicolumn{5}{c}{Cesium} \\
RPA & -3.041 & 11.965 & -0.0249 & 8.899 \\
BO\tablenotemark[1]  & -3.358 & 12.454 & -0.0242 & 9.072 \\
$\Delta$& -0.316 &  0.489 & -0.001 & 0.173 \\
$\Delta$ (\%) & -3.5\% &  5.4\% &    0.0\% & 2.0\% \\

\multicolumn{5}{c}{Rubidium} \\
RPA &-0.408 &  1.756 &-0.003 & 1.345 \\
BO\tablenotemark[1]  &-0.463 &  1.866 &-0.003 & 1.400 \\
$\Delta$&-0.055 &  0.011 & 0.000 & 0.055 \\
$\Delta$ (\%) & -3.9\% &  7.8\% &    0.0\% & 4.0\% \\
\end{tabular}
\tablenotetext[1]{Brueckner orbitals with core polarization.}
\end{ruledtabular}
\end{table}

\paragraph{Breit interaction.} In was demonstrated in~\cite{D-Breit}
that Breit interaction gives significant contribution to the PNC in
many-electron atoms. The contribution of Breit interaction is about
-0.6\% to the PNC amplitude in Cs~\cite{D-Breit,DzBreit1} (see also
Table~\ref{t:pnc}). To calculate Breit correction 
in Rb we use the same approach as in our previous
works~\cite{DzBreit1,DzBreit2}. The Breit Hamiltonian includes magnetic and
retardation terms. The Coulomb interaction everywhere in the
calculations is replaced by the sum of Coulomb and Breit terms $V
\rightarrow V_C + V_B$. Second-order correlation correction operator
$\hat \Sigma$ is used to calculate Brueckner orbitals. The correction
is found as a difference between two results for Eq.~(\ref{amp}), one
with the  Breit interaction  included and another when it is not included.   
The resulting Breit correction is about -0.4\% of the PNC amplitude in
Rb. Its relative value is about 1.5 times smaller than in
Cs. Therefore, the uncertainty associated with this correction is also
smaller for Rb than for Cs.

\paragraph{QED corrections.}

The quantum electrodynamics corrections to the $E_{PNC}$ for rubidium
are calculated using the sum-over-states method.  The sum which needs
to be evaluated is 
\begin{align}
E_{\rm PNC} = \sum_n
& \Bigg[\frac{\langle{6s|\tilde{d}}|{np_{1/2}}\rangle
  \langle{np_{1/2}|\tilde H_{W}}|{5s}\rangle}{E_{5s}-E_{np}}  
        \notag \\ 
&+\frac{\langle 6s|\tilde H_{W}|np_{1/2}\rangle \langle
  np_{1/2}|\tilde{d}|{5s}\rangle}{E_{6s}-E_{np}}\Bigg].  
	\label{eq:qed}
\end{align}
Here tilde means that core polarization is taken into account
(e.g. $\tilde d = \hat H_{E1} + \delta \hat V_{E1}$). Correlations are
taken into account by using Brueckner orbitals for all $ns$ and
$np_{1/2}$ states.
 
The QED corrections to the PNC amplitude can be seen to arise from
three different sources.  There are  the corrections to the $E1$
dipole matrix elements ($\tilde{d}$), weak-interaction matrix
elements ($\tilde H_{W}$), and the energy denominators. Corrections to the
weak matrix elements have been considered previously
\cite{kuchiev2003radiative,milstein2003calculation} (see also  
\cite{shabaev2005qed}).  From these works, we determine the QED
contribution to the PNC amplitude coming from corrections to the weak
matrix elements elements to be $-0.30(2)\%$. 

For the corrections coming from the energy denominators and dipole
amplitudes, we use the ``radiative potential'' method proposed in
\cite{Gingesrad}.  By calculating the dominating terms in equation
(\ref{eq:qed}) both with and without QED corrections, we determine the
correction coming from the energy denominators to be $-0.25\%$ and
from the dipole amplitudes to be $+0.31\%$ giving a combined shift of
$+0.06(3)\%$ for dipoles and energies. 

Therefore we find the total QED shift to the $5s$-$6s$ PNC amplitude
in rubidium to be $-0.24(4)\%$.  As expected, the Rb result is smaller
than that in Cs atom. More importantly, omitted higher order
corrections in $Z \alpha$ should be much smaller in Rb. 

\paragraph{Neutron skin.}

The neutron skin correction to the PNC amplitude is due to the fact
that nuclear density in the weak interaction Hamiltonian (\ref{e1}) is
not the same as nuclear charge distribution. This density is dominated
by neutrons and if the neutron distribution radius differs from the
radius of the proton distribution this would lead to a correction to the
PNC amplitude. It was found from an analysis of data for
antiprotonic atoms~\cite{RMS} that the root mean square radius of the
proton and neutron distributions differ by
\begin{equation}
  \Delta r_{np} = (-0.04 \pm 0.03) + (1.01 \pm 0.15)\frac{N-Z}{A} \
  {\rm fm}.
\label{eq:rnp}
\end{equation}
Using this data to correct nuclear density in (\ref{e1}) and
recalculating the PNC amplitude leads to -0.2\% correction for Cs and
-0.06\% correction for Rb (see Table~\ref{t:pnc}). Here again the
correction is much smaller for Rb than for Cs leading to smaller
uncertainty in the PNC amplitude. 

Combining all corrections we obtain the final value of the $5s$ -
$6s$ nuclear-spin-independent PNC amplitude in $^{87}$Rb:
\begin{equation}
|E_{\rm PNC}| = 1.390(7) \times 10^{-12} ea_B (-Q_W/N).
\label{eq:epncfinal}
\end{equation}
and for $^{85}$Rb:
\begin{equation}
|E_{\rm PNC}| = 1.333(7) \times 10^{-12} ea_B (-Q_W/N).
\label{eq:epncfinal}
\end{equation}
We assume the 0.5\% uncertainty as it has been discussed above.

The nuclear spin-dependent PNC amplitudes for transitions between
different hyperfine structure components of the $5s$ and $6s$ states
of $^{85}$Rb and $^{87}$Rb are presented in Table~\ref{t:sdpnc}.

\begin{table}
\caption{PNC amplitudes ($z$-components) for the $|5s,F_1
  \rangle \rightarrow  |6s,F_2\rangle$ transitions in
  $^{85}$Rb and $^{87}$Rb. Units: $10^{-11} iea_0$.} 
\label{t:sdpnc}
\begin{ruledtabular}
\begin{tabular}{ccccc r}
Isotope & $Q_W$ & $I$ & $F_1$ & $F_2$ & \multicolumn{1}{c}{PNC amplitude} \\
\hline
$^{85}$Rb & -44.76 &2.5 & 2 & 2 & $ 0.091(1+ 0.153\varkappa)$ \\
         &        &    & 2 & 3 & $ -0.102(1-0.206\varkappa)$ \\
         &        &    & 3 & 2 & $ -0.102(1+ 0.250\varkappa)$ \\
         &        &    & 3 & 3 & $ -0.137(1-0.109\varkappa)$ \\
$^{87}$Rb& -46.74 & 1.5 & 1 & 1 & $  0.071(1+ 0.105\varkappa)$ \\
         &       &     & 1 & 2 & $ -0.123(1-0.125\varkappa)$ \\
         &       &     & 2 & 1 & $ -0.123(1+ 0.167\varkappa)$ \\
         &       &     & 2 & 2 & $ -0.143(1-0.063\varkappa)$ \\
\end{tabular}
\end{ruledtabular}
\end{table}

\paragraph{Acknowledgments.}

The work was supported in part by the Australian Research Council.

\end{document}